\documentclass[conf]{new-aiaa}
\usepackage[utf8]{inputenc}
\usepackage{listings}
\usepackage{xcolor}
\usepackage{subfigure}

\definecolor{codegreen}{rgb}{0,0.6,0}
\definecolor{codegray}{rgb}{0.5,0.5,0.5}
\definecolor{codepurple}{rgb}{0.58,0,0.82}
\definecolor{backcolour}{rgb}{0.95,0.95,0.92}

\lstdefinestyle{mystyle}{
    backgroundcolor=\color{backcolour},   
    commentstyle=\color{codegreen},
    keywordstyle=\color{magenta},
    numberstyle=\tiny\color{codegray},
    stringstyle=\color{codepurple},
    basicstyle=\ttfamily\footnotesize,
    breakatwhitespace=false,         
    breaklines=true,                 
    captionpos=b,                    
    keepspaces=true,                 
    numbersep=5pt,                  
    showspaces=false,                
    showstringspaces=false,
    showtabs=false,                  
    tabsize=2
}

\usepackage{graphicx}
\usepackage{amsmath}
\usepackage[version=4]{mhchem}
\usepackage{siunitx}
\usepackage{longtable,tabularx}
\title{Deploying deep learning in OpenFOAM with TensorFlow}

\author{Romit Maulik\footnote{AIAA Member.}, Himanshu Sharma\footnote{Now at Pacific Northwest National Laboratory, Richland WA-99354, USA}}
\affil{Argonne Leadership Computing Facility, Argonne National Laboratory, Lemont IL-60439, USA}
\author{Saumil Patel}
\affil{Computational Science Division, Argonne National Laboratory, Lemont, IL-60439, USA}
\author{Bethany Lusch, Elise Jennings \footnote{Now at Irish Centre for High-End Computing at NUI Galway}}
\affil{Argonne Leadership Computing Facility, Argonne National Laboratory, Lemont IL-60439, USA}

\begin{document}

\maketitle

\begin{abstract}
We outline the development of a data science module within OpenFOAM which allows for the in-situ deployment of trained deep learning architectures for general-purpose predictive tasks. This module is constructed with the TensorFlow C API and is integrated into OpenFOAM as an application that may be linked at run time. Notably, our formulation precludes any restrictions related to the \emph{type} of neural network architecture (i.e., convolutional, fully-connected, etc.). This allows for potential studies of complicated neural architectures for practical CFD problems. In addition, the proposed module outlines a path towards an open-source, unified and transparent framework for computational fluid dynamics and machine learning.
\end{abstract}

\section{Nomenclature}

{\renewcommand\arraystretch{1.0}
\noindent\begin{longtable*}{@{}l @{\quad=\quad} l@{}}
$C_s$ & Dynamic Smagorinsky coefficient \\
$h$  & Backward facing step-height \\
$\nu_t$ & Turbulent eddy-viscosity \\
$Re_\tau$ & Friction Reynolds number \\
$R_{ij}$ & Reynolds stress tensor
\end{longtable*}}

\section{Introduction}

In recent times, physics-informed machine learning algorithms have generated a lot of interest for computational fluid dynamics (CFD) applications. These algorithms have been applied for wide variety of tasks such as closure modeling \cite{beck2019deep,maulik2019sub,singh2017machine,gamahara2017searching,ling2016reynolds,ling2015evaluation,matai2019zonal,taghizadeh2020turbulence,sotgiu2019towards,wu2018physics,wu2019physics,xiao2016quantifying,zhang2019recent}, control \cite{muller1999application,lee1997application,gautier2015closed,duriez2017machine,raibaudo2020machine}, surrogate modeling \cite{zhu2019machine,trehan2017error,san2018extreme,san2018machine,renganathan2020machine,qian2020lift,lee2020model,hasegawa2020machine,mohan2018deep,mohan2019compressed,maulik2020non,maulik2020time,fukami2020convolutional,maulikrecurrent2020}, inverse problems \cite{raissi2020hidden,raissi2019physics,gao2020bi,sun2020physics,sun2020surrogate}, uncertainty quantification \cite{kawai2014kriging,geneva2019quantifying,maulik2020probabilistic}, data assimilation \cite{gao2020bi,tang2020deep,casas2020reduced,pawar2020data} and super-resolution \cite{fukami2019super,liu2020deep}. These studies have demonstrated that the ability of modern machine learning algorithms to learn complicated nonlinear relationships may be leveraged for improving accuracy of quantities of interest as well as significant reductions in computational costs. Indeed, these studies have generated exciting results in a wide range of research thrusts within computational physics such as for novel turbulence models, shock capturing methods, shape optimization, adaptive mesh refinement strategies, interface tracking algorithms. Exhaustive reviews of machine learning methods for fluid dynamics can be found in Brunton et al. \cite{brunton2020machine} and Fukami et al. \cite{fukami2020assessment}, and a review of machine learning opportunities for turbulence may be found in Duraisamy et al. \cite{duraisamy2019turbulence}

Our present study is motivated by this explosion of data-driven algorithm development. Specifically, we seek to address the lack of a coherent framework for reproducible data-driven algorithm development and testing. Our goal, for this study, is to propose an in-situ coupling mechanism between data science and simulation that may be utilized for a large variety of simulation tasks as well as machine learning strategies. Crucially, we also wish to build on an well-established CFD package that has been tested for a wide variety of problems deployed both serially and in parallel. This would allow for the easy integration of machine learning into a previously established meshing, simulation and visualization assembly line.

We choose OpenFOAM \cite{weller1998tensorial}, an open-source general-purpose CFD software under active development, as our simulation framework. For our data-science capability, we choose TensorFlow 1.15 \cite{tensorflow2015-whitepaper}, a well-known machine learning library that allows for the development of data-driven techniques as simple as linear, logistic and polynomial regression or as complicated as fully connected and convolutional neural networks, regular and variational autoencoders and generative adversarial networks. This article serves as an introduction as well as a tutorial to the proposed coupling mechanism. The coupling is introduced by way of a surrogate prediction task for the Reynolds-averaged Navier-Stokes (RANS) turbulence eddy viscosity for a canonical backward facing step problem (with greater detail available in \cite{maulik2019accelerating}). We also demonstrate its viability for competitive compute times by deploying a deep learning surrogate to the dynamic Smagorinsky model \cite{germano1991dynamic} for a turbulent channel flow at $Re_\tau=180$. We shall introduce our coupling mechanism through a step by step demonstration that can be implemented by an OpenFOAM user independently. In addition, all of our source code and data are available publicly \footnote{https://github.com/argonne-lcf/TensorFlowFoam}.

\section{Literature review}

Despite the massive popularity of machine learning algorithm development for various tasks, there have been few open-source frameworks that have successfully allowed for direct embedding of general-purpose machine learning algorithms within simulation codes in an easy-to-reproduce manner. One such example is that of Geneva and Zabaras \cite{geneva2019quantifying} who embed a neural network model from PyTorch into OpenFOAM 4.1. However, this procedure requires the installation of additional software packages such as ONNX and Caffe2 that may cause issues with dependencies. In addition, Caffe2 is deprecated (to be subsumed into PyTorch) and future incorporation of PyTorch models into OpenFOAM through this route is unclear. Another framework under active development is the Fortran-Keras Bridge \cite{ott2020fortran} that successfully couples densely connected neural networks to Fortran simulation codes. However, this framework is yet to support more complicated architectures such as convolutional neural networks and development revolves around the programming of neural network subroutines in Fortran before a Keras model can be imported. In contrast, we utilize TensorFlow to represent a generic deep learning architecture as a graph on disk which is imported into OpenFOAM during runtime.

\section{The coupling mechanism}

In the following section, we outline the procedure for using the TensorFlow C API in OpenFOAM. In addition we detail how to export a trained deep learning model from TensorFlow in Python to disk as a protobuffer (where all of the trainable parameters and operations of the deep learning framework are fixed). Finally, we outline how to load this protobuffer during the solution process and interface it with OpenFOAM data structures.

\subsection{Exporting model from Python}

A generic TensorFlow model may be defined in Python by using the Keras API. A simple fully-connected network is defined as follows: 

\begin{lstlisting}[language=Python,style=mystyle,caption=Fully connected network definition using the Keras API of TensorFlow 1.15 in Python 3.6.8.]
from tensorflow import keras
def get_model(num_inputs,num_outputs,num_layers,num_neurons):
    '''
    num_inputs  : Number of model inputs
    num_outputs : Number of model outputs
    num_layers  : Number of hidden layers
    num_neurons : Number of neurons in hidden layers
    Returns: TensorFlow model
    '''
    
    # Input layer
    ph_input = keras.Input(shape=(num_inputs,),name='input_placeholder')
    
    # Hidden layers
    hidden_layer = keras.layers.Dense(num_neurons,activation='tanh')(ph_input)

    for layer in range(num_layers):
        hidden_layer = keras.layers.Dense(num_neurons,activation='tanh')(hidden_layer)

    # Output layer
    output = keras.layers.Dense(num_outputs,activation='linear',name='output_value')(hidden_layer)

    model = keras.Model(inputs=[ph_input],outputs=[output])
    
    # Optimizer
    my_adam = keras.optimizers.Adam(lr=0.001, decay=0.0)

    # Compilation
    model.compile(optimizer=my_adam,loss={'output_value': 'mean_squared_error'})

    return model
\end{lstlisting}

The abstractions of this API allow for significant complexity in model development. After this model has been defined and trained, the first step for exporting the model to a non-pythonic environment requires a function to freeze the model weights. This is achieved through the following method: 

\begin{lstlisting}[language=Python,style=mystyle,caption=Freezing trained neural network model to protobuffer format.]
def freeze_session(session, keep_var_names=None, output_names=None, clear_devices=True):
    """
    Freezes the state of a session into a pruned computation graph. Creates a new computation graph where variable nodes are replaced by constants taking their current value in the session. 
    session: The TensorFlow session to be frozen.
    keep_var_names: A list of variable names that should not be frozen, or None to freeze all the variables in the graph.
    output_names: Names of the relevant graph outputs.
    clear_devices: Remove the device directives from the graph for better portability.
    Returns: The frozen graph definition.
    """
    graph = session.graph
    with graph.as_default():
        freeze_var_names = list(set(v.op.name for v in tf.global_variables()).difference(keep_var_names or []))
        output_names = output_names or []
        output_names += [v.op.name for v in tf.global_variables()]
        input_graph_def = graph.as_graph_def()
        if clear_devices:
            for node in input_graph_def.node:
                node.device = ""
        frozen_graph = tf.graph_util.convert_variables_to_constants(
            session, input_graph_def, output_names, freeze_var_names)
        return frozen_graph

# Save the graph to disk
frozen_graph = freeze_session(keras.backend.get_session(),output_names=[out.op.name for out in model.outputs])
tf.train.write_graph(frozen_graph, './', 'ML_Model.pb', as_text=False)
\end{lstlisting}

The model (in the protobuffer format) can now be imported using the C API.

\subsection{The TensorFlow C API}

Next, we present the procedure to call the TensorFlow C API for loading a graph and performing an inference within a general C++ code. This is accomplished as follows:
\begin{lstlisting}[language=C++,style=mystyle,caption=Loading model saved in protobuffer]
// TensorFlow C API header
#include <tensorflow/c/c_api.h> 
// Function to load a graph from protobuffer
TF_Graph* LoadGraph(const char* graphPath) {
  if (graphPath == nullptr) {
    return nullptr;
  }

  TF_Buffer* buffer = ReadBufferFromFile(graphPath);
  if (buffer == nullptr) {
    return nullptr;
  }

  TF_Graph* graph = TF_NewGraph();
  TF_Status* status = TF_NewStatus();
  TF_ImportGraphDefOptions* opts = TF_NewImportGraphDefOptions();

  TF_GraphImportGraphDef(graph, buffer, opts, status);
  TF_DeleteImportGraphDefOptions(opts);
  TF_DeleteBuffer(buffer);

  if (TF_GetCode(status) != TF_OK) {
    TF_DeleteGraph(graph);
    graph = nullptr;
  }

  TF_DeleteStatus(status);

  return graph;
}

// Load graph from disk
graph_ = LoadGraph("./ML_Model.pb");

// Input operation
input_ph_ = {TF_GraphOperationByName(graph_, "input_placeholder"), 0};

// Output operation
output_ = {TF_GraphOperationByName(graph_, "output_value/BiasAdd"), 0};
\end{lstlisting}

Note how the names of the operations defined in the Python model are required to identify operations in C++. An inference using this loaded graph may be performed using

\begin{lstlisting}[language=C++,style=mystyle,caption=Inference using TF C API in C++]
TF_Tensor* CreateTensor(TF_DataType data_type,
                        const std::int64_t* dims, std::size_t num_dims,
                        const void* data, std::size_t len) {
  if (dims == nullptr) {
    return nullptr;
  }

  TF_Tensor* tensor = TF_AllocateTensor(data_type, dims, static_cast<int>(num_dims), len);
  if (tensor == nullptr) {
    return nullptr;
  }

  void* tensor_data = TF_TensorData(tensor);
  if (tensor_data == nullptr) {
    TF_DeleteTensor(tensor);
    return nullptr;
  }

  if (data != nullptr) {
    std::memcpy(tensor_data, data, std::min(len, TF_TensorByteSize(tensor)));
  }

  return tensor;
}

void DeleteTensor(TF_Tensor* tensor) {
  if (tensor != nullptr) {
    TF_DeleteTensor(tensor);
  }
}

void DeleteSession(TF_Session* session) {
  TF_Status* status = TF_NewStatus();
  TF_CloseSession(session, status);
  if (TF_GetCode(status) != TF_OK) {
    TF_CloseSession(session, status);
  }
  TF_DeleteSession(session, status);
  if (TF_GetCode(status) != TF_OK) {
    TF_DeleteSession(session, status);
  }
  TF_DeleteStatus(status);
}

TF_Status* status_ = TF_NewStatus();
TF_SessionOptions* options_ = TF_NewSessionOptions();
TF_Session* sess_ = TF_NewSession(graph_, options_, status_);

TF_Tensor* output_tensor_ = nullptr;
TF_Tensor* input_tensor_ = CreateTensor(TF_FLOAT,
                                      input_dims.data(), input_dims.size(),
                                      &input_vals, num_cells*num_inputs*sizeof(float));
    
// Arrays of tensors
TF_Tensor* input_tensors_[1] = {input_tensor_};
TF_Tensor* output_tensors_[1] = {output_tensor_};
// Arrays of operations
TF_Output inputs[1] = {input_ph_};
TF_Output outputs[1] = {output_};
    
TF_SessionRun(sess_,
            nullptr, // Run options.
            inputs, input_tensors_, 1, // Input tensor ops, input tensor values, number of inputs.
            outputs, output_tensors_, 1, // Output tensor ops, output tensor values, number of outputs.
            nullptr, 0, // Target operations, number of targets.
            nullptr, // Run metadata.
            status_ // Output status.
            );

// Cast output from TF C API to float*
const auto data = static_cast<float*>(TF_TensorData(output_tensors_[0]));

DeleteTensor(input_tensor_);
DeleteTensor(output_tensor_);
TF_DeleteSessionOptions(options_);
TF_DeleteStatus(status_);
DeleteSession(sess_);
\end{lstlisting}
where \texttt{input\_vals} and \texttt{input\_dims} are the data and the dimensions, respectively, for the input data. 

\subsection{Compiling an OpenFOAM turbulence model that calls TensorFlow}

Once we have established how to deploy a trained machine learning model in C++, we may call for an inference within OpenFOAM. For the purpose of demonstration, we show how a new turbulence model library may be compiled while linking to the TensorFlow C Libraries. To start, we must give OpenFOAM the path to the TensorFlow C API in the \lstinline[columns=fixed,style=mystyle]{Make/Options} file as follows
\begin{lstlisting}[language=bash,style=mystyle,caption=Make/Options for OpenFOAM and TensorFlow linkage]
EXE_INC = \
    -I$(LIB_SRC)/TurbulenceModels/turbulenceModels/lnInclude \
    -I$(LIB_SRC)/TurbulenceModels/incompressible/lnInclude \
    -I$(LIB_SRC)/transportModels \
    -I$(LIB_SRC)/finiteVolume/lnInclude \
    -I$(LIB_SRC)/meshTools/lnInclude \
    -I/path/to/tensorflow/include \
LIB_LIBS = \
    -L/path/to/tensorflow/lib \
    -lincompressibleTransportModels \
    -lturbulenceModels \
    -lfiniteVolume \
    -lmeshTools \
    -ltensorflow \
    -lstdc++
\end{lstlisting}
One may now compile a new turbulence model using \texttt{wmake}. 

\section{Demonstration}

In the following section, we shall outline some results from deploying deep learning inference within OpenFOAM 5.0 simulations for canonical turbulence problems. We shall also demonstrate the potential benefits of using ML within scientific computing with results that exhibit reduced time to solution. We note that all experiments within this section were carried out on an 8th-generation Intel CoreI7 processor with a clockspeed of approximately 1.90GHz and on an Ubuntu 18.04 operating system.

\subsection{Science driver 1: Surrogate modeling for steady-state turbulent eddy-viscosity}

In this section, we demonstrate how the direct prediction of the steady-state turbulent eddy-viscosity (as a function of initial conditions) can accelerate RANS simulations. Our framework is outlined in greater detail in \cite{maulik2019accelerating}. Briefly, the initial conditions for a RANS simulation (coming from a low-fidelity potential solution) on an arbitrary mesh can be used to predict an approximation to the steady-state Spalart-Allmaras turbulent eddy-viscosity \cite{spalart1992one} via a neural network. Our training data is obtained by performing multiple RANS simulations using the Spalart-Allmaras model on different geometries. In this experiment, the different geometries are all backward facing steps with varying step heights ($h$).

\begin{figure}
    \centering
    \mbox{
    \subfigure[$\nu_t$ predicted]{\includegraphics[trim={0 6cm 0 0},clip,width=0.48\textwidth]{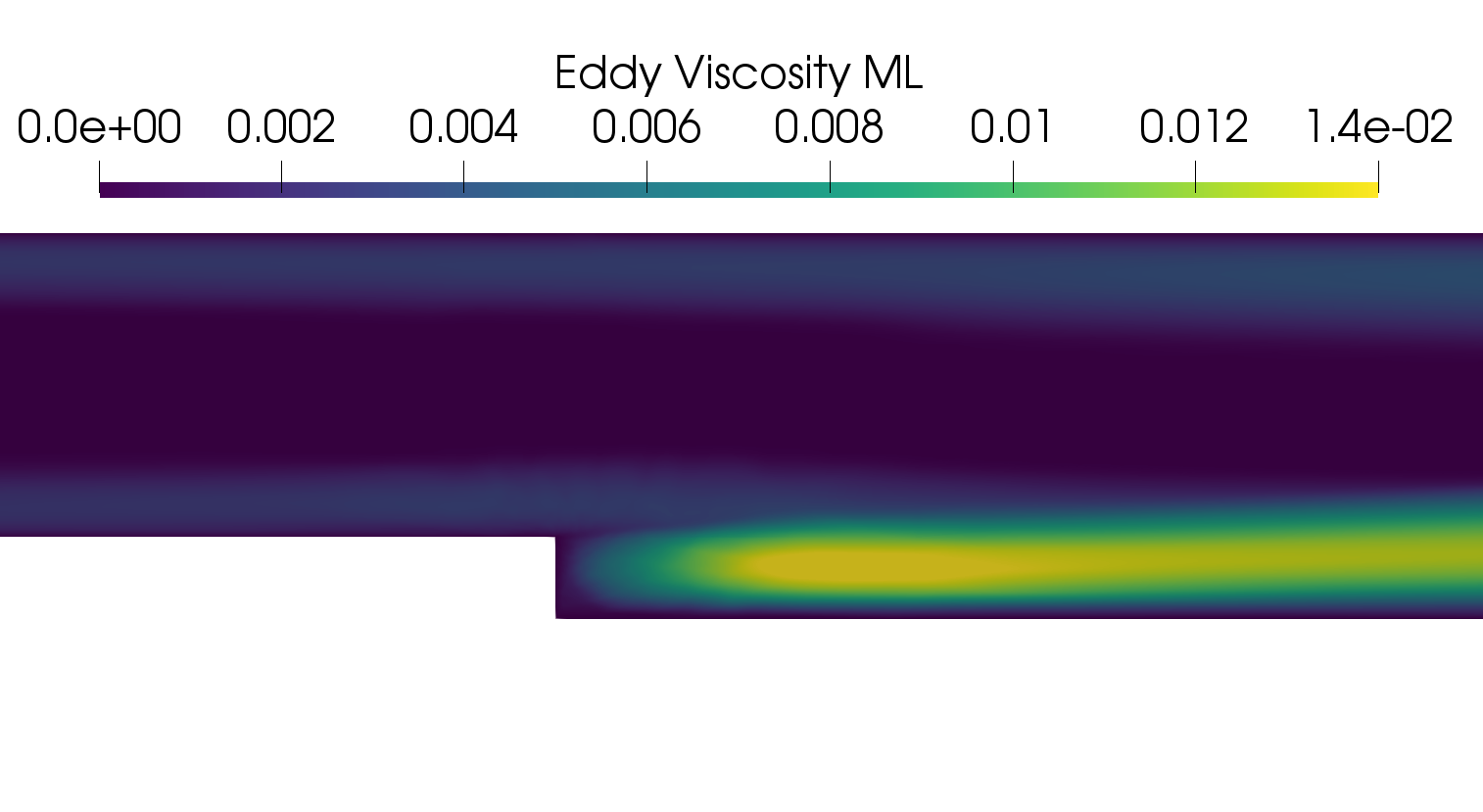}}
    \subfigure[$|U|$ predicted]{\includegraphics[trim={0 6cm 0 0},clip,width=0.48\textwidth]{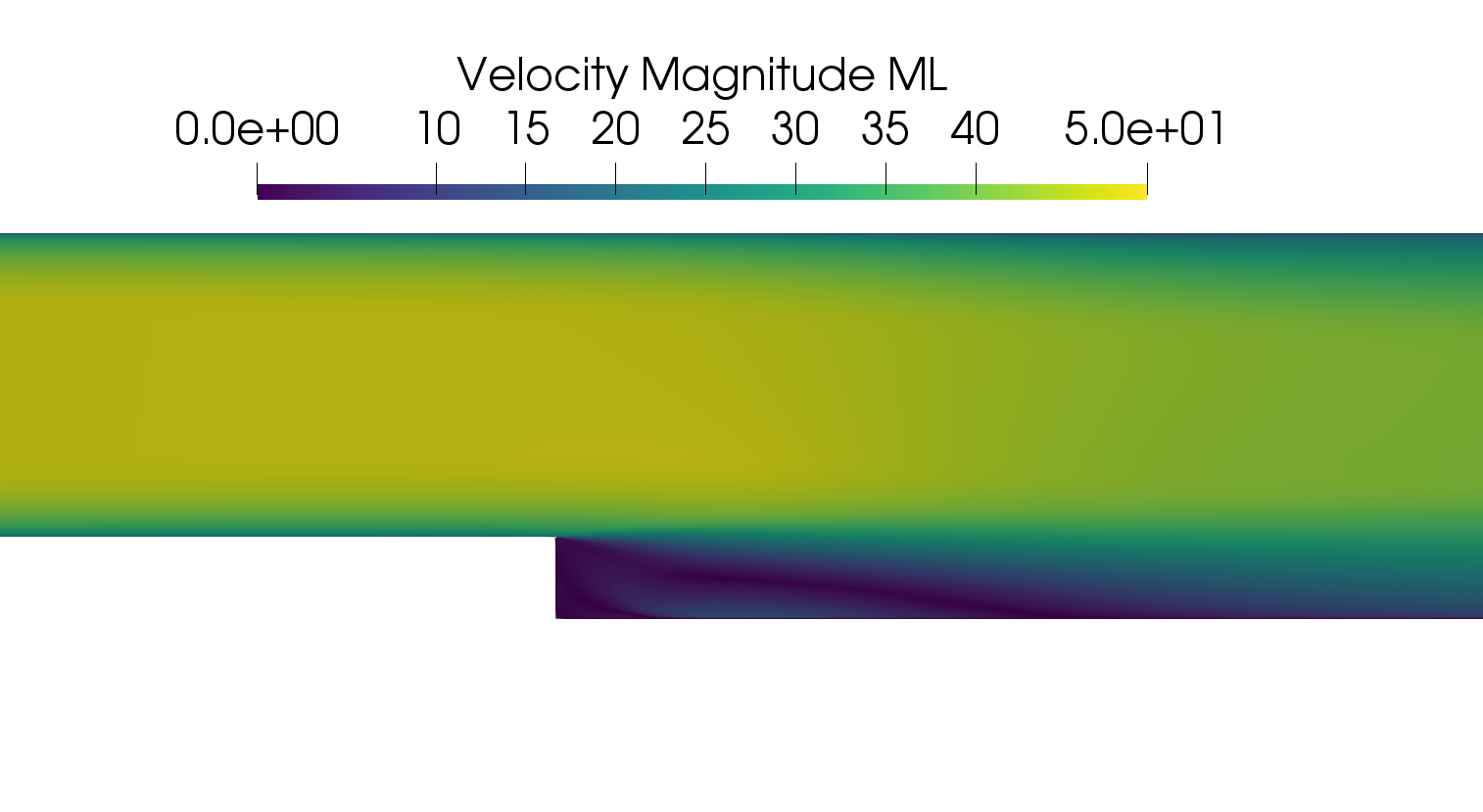}}
    } \\
    \mbox{
    \subfigure[$\nu_t$ true]{\includegraphics[trim={0 6cm 0 0},clip,width=0.48\textwidth]{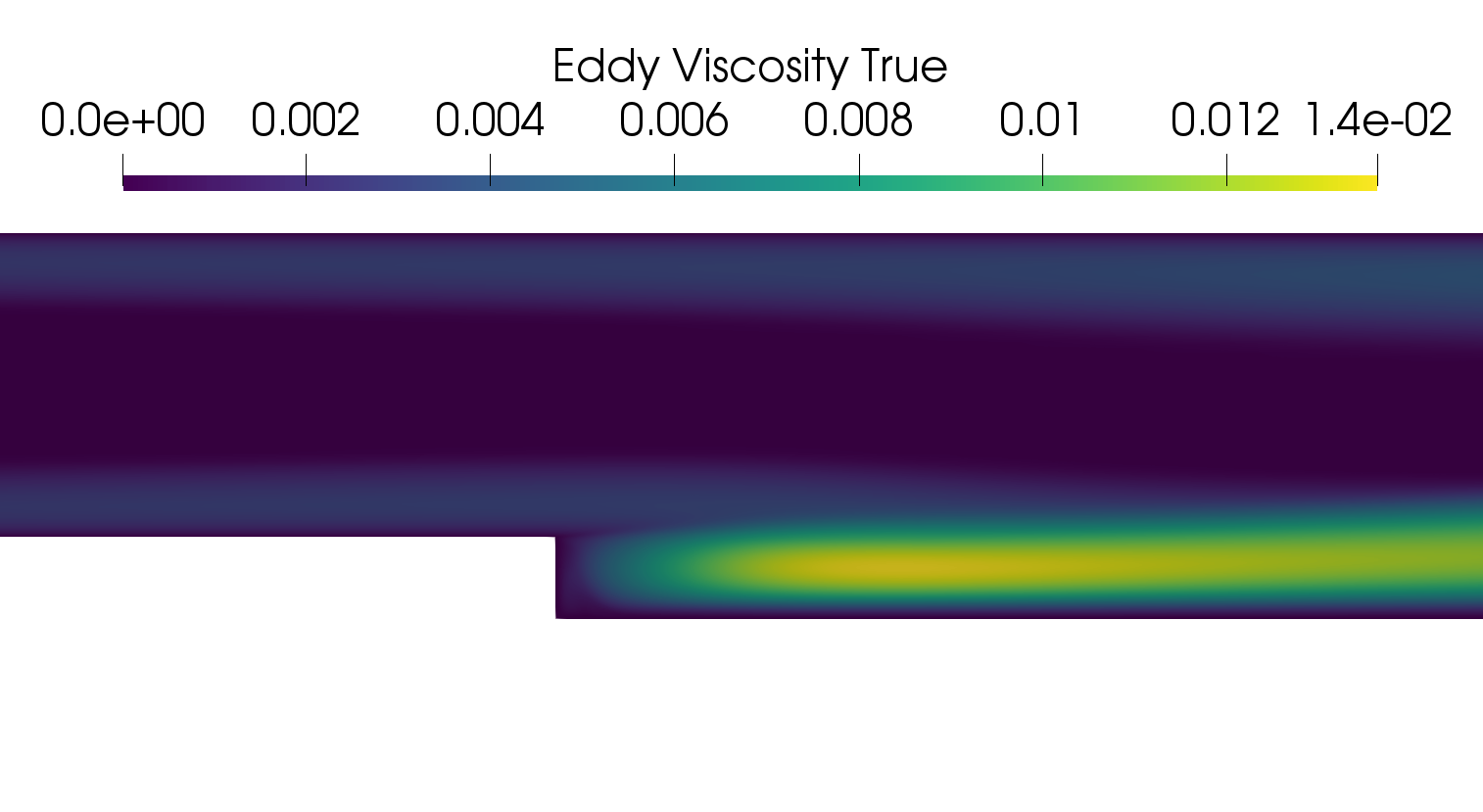}}
    \subfigure[$|U|$ true]{\includegraphics[trim={0 6cm 0 0},clip,width=0.48\textwidth]{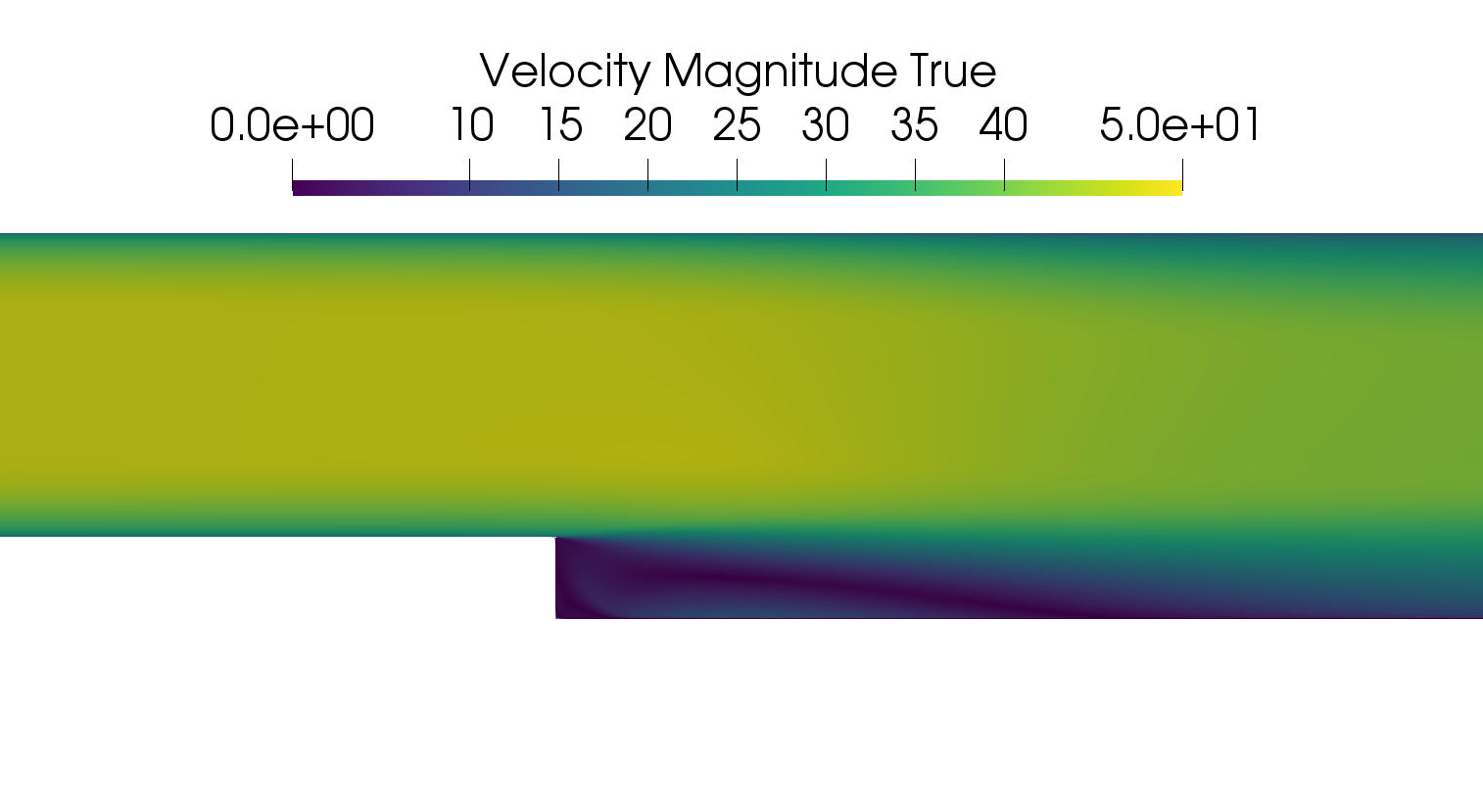}}
    }
    \caption{Contour plots for a backward facing step. Note that the training of the ML surrogate did not include data for the shown step height.}
    \label{RANS_1}
\end{figure}

\begin{figure}
    \centering
    \mbox{
    \includegraphics[width=\textwidth]{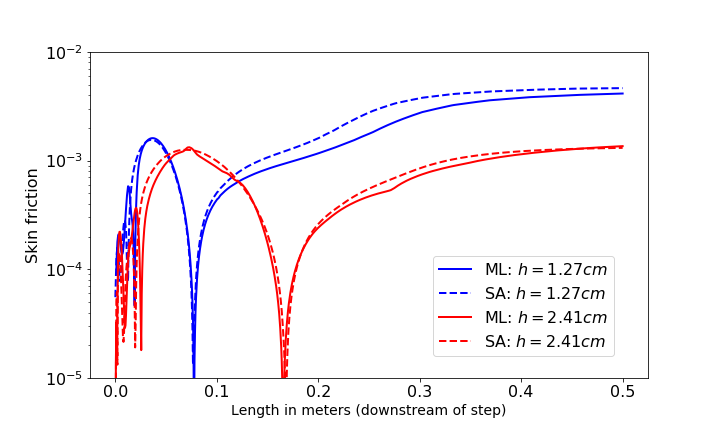}
    }
    \caption{Skin-friction coefficient predictions downstream of the step when deploying the ML framework on two test step heights. The ML framework is seen to introduce some errors. Downstream skin friction inaccuracies suggest adaptive sampling may be necessary for improved generalization across geometries. However, the reattachment point is recovered well.}
    \label{RANS_2}
\end{figure}

Once trained, the steady-state eddy-viscosity emulator may be used at the start of the simulation (by observing the initial conditions) following which solely the pressure and velocity equations need to be solved to convergence. We outline results from one such experiment, where the geometry is `unseen', in Figure \ref{RANS_1}. The time-to-solution of the proposed framework is significantly reduced, albeit at the cost of some increase in error, as shown in Figure \ref{RANS_2}. Potential extensions to such a framework include trying to bypass high fidelity turbulence models (two-equation) or generating models from temporally averaged DNS data. 

\subsection{Science driver 2: Surrogate modeling of dynamic Smagorinsky coefficient}

In this section, we show how a deep learning framework to predict the Smagorinsky coefficient (dynamically) may be used within OpenFOAM. Our test case is given by a turbulent channel flow at Re$_\tau=395$. For a proof of concept, we use our data-driven model to predict the dynamic Smagorinsky coefficient $C_s$, given instantaneous measurements of the strain-rate tensor as well as the three velocity components at each cell center of the mesh. For preliminary results, our training data is generated from Dynamic Smagorinsky itself (i.e., with a standard scale-similarity based least-squares calculation following test-filtering). The goal, for this exercise, is to see if a deep neural network, despite a far greater number of floating point operations, can efficiently bypass the dynamic Smagorinsky framework. Success in this regard could lead to the development of closures from data obtained on higher fidelity meshes as well as from DNS while retaining efficient inference speeds on the coarser meshes.

\begin{figure}
    \centering
    \mbox{
    \subfigure[$R_{11}$]{\includegraphics[width=0.48\textwidth]{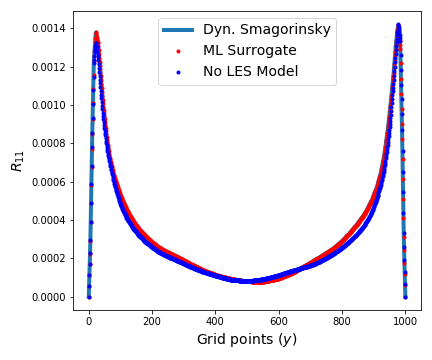}}
    \subfigure[$R_{22}$]{\includegraphics[width=0.48\textwidth]{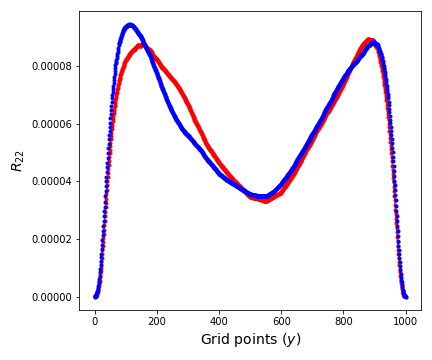}}
    } \\
    \mbox{
    \subfigure[$R_{33}$]{\includegraphics[width=0.48\textwidth]{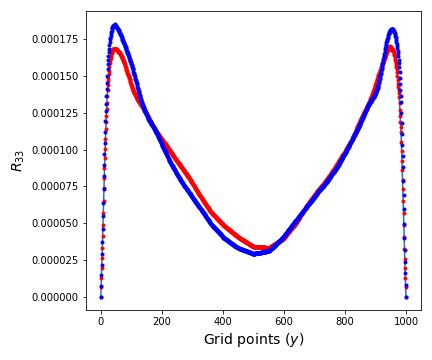}}
    \subfigure[$R_{12}$]{\includegraphics[width=0.48\textwidth]{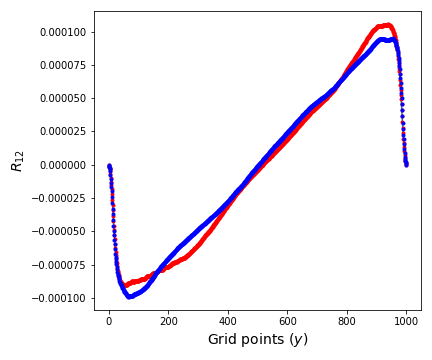}}
    } \\
    \caption{Ensemble averaged Reynolds stresses for channel flow at $Re_\tau=395$, where  $R_{ij} = \langle u_i' u_j' \rangle - \langle u_i'\rangle \langle u_j'\rangle$. It is seen that the the surrogate model using a neural network for Smagorinsky coefficient calculation is able to recreate the Dynamic Smagorinsky solution with comparable times to solution. The ``No LES Model" curve indicates a computation without the use of any turbulence modeling and therefore causes deviation from the that obtained by Dynamic Smagorinsky.}
    \label{fig:les}
\end{figure}

We show a preliminary analysis in Figure \ref{fig:les} for the ensemble-averaged Reynolds stresses given by
\begin{align}
    R_{ij} = \langle u_i' u_j' \rangle - \langle u_i'\rangle \langle u_j'\rangle
\end{align}
where the angled-brackets imply time-averaging and $u_i^{'} = u_i - \langle u_i \rangle$ is the temporal fluctuation. It is observed that the ML surrogate recovers performance similar to Dynamic Smagorinsky without the use of test-filtering. The compute times to solution for both cases on one node (identical to the node used for the RANS experiments previously) were 1348 seconds for Dynamic Smagorinsky and 1297 seconds for the ML surrogate assuming a physical solution time of 1000 seconds. This is promising since the neural network requires far more floating point operations than the standard dynamic Smagorinsky coefficient calculation. 

\section{Conclusion and Future Work}

Ongoing work related to the content introduced in this article is focused on the assessment of ML inference at scale. It is important to assess how in-situ ML inference frameworks may affect CFD workflows optimized for distributed parallelism. In addition, we have also focused our efforts on developing a viable `training-online' strategy (also within OpenFOAM) where a neural architecture may be trained while a simulation is actually running. As stated previously, the presence of these capabilities in a common framework will allow for greater transparency and reproducibility of data-driven algorithms for scientific computing.

\section*{Acknowledgments}

This material is based upon work supported by the U.S. Department of Energy (DOE), Office of Science, Office of Advanced Scientific Computing Research, under Contract DE-AC02-06CH11357. This research   was funded in part and used resources of the Argonne Leadership Computing Facility, which is a DOE Office of Science User Facility supported under Contract DE-AC02-06CH11357. This paper describes objective technical results and analysis. HS acknowledges support from the ALCF Director's Discretionary (DD) program for CFDML project. Any subjective views or opinions that might be expressed in the paper do not necessarily represent the views of the U.S. DOE or the United States Government.

\bibliography{sample}

\end{document}